\documentclass[
               twocolumn,
               noshowpacs,          
               nopreprintnumbers,     
               aps,                 
               prl,                 
               a4paper,             
               superscriptaddress,  
               nofootinbib,         
               tightenlines,        
               floats,floatfix      
               ]{revtex4}

\usepackage{amsmath, amsfonts, amsthm, amssymb, graphicx,vector}
\usepackage{epstopdf}
 \usepackage{slashed}
\usepackage{feynmp}

\def\be{\begin{equation}}
\def\ee{\end{equation}}
\def\ba{\begin{eqnarray}}
\def\ea{\end{eqnarray}}

\def\bdm{\begin{displaymath}}
\def\edm{\end{displaymath}}
\def\la{~\mbox{\raisebox{-.6ex}{$\stackrel{<}{\sim}$}}~}
\def\ga{~\mbox{\raisebox{-.6ex}{$\stackrel{>}{\sim}$}}~}
\def\bq{\begin{quote}}
\def\eq{\end{quote}}
\def\la{~\mbox{\raisebox{-.6ex}{$\stackrel{<}{\sim}$}}~}
\def\ga{~\mbox{\raisebox{-.6ex}{$\stackrel{>}{\sim}$}}~}

\def\del{\partial}
\def\ltap{\ \raise.3ex\hbox{$<$\kern-.75em\lower1ex\hbox{$\sim$}}\ }
\def\gtap{\ \raise.3ex\hbox{$>$\kern-.75em\lower1ex\hbox{$\sim$}}\ }
\def\gl{\ \raise.5ex\hbox{$>$}\kern-.8em\lower.5ex\hbox{$<$}\ }
\def\roughly#1{\raise.3ex\hbox{$#1$\kern-.75em\lower1ex\hbox{$\sim$}}}

 at 10truept

\def\k{{\uvec{k}}}

\newcommand{\beq}{\begin{equation}}
\newcommand{\eeq}{\end{equation}}
\newcommand{\bea}{\begin{eqnarray}}
\newcommand{\eea}{\end{eqnarray}}
\newcommand{\beqa}{\begin{eqnarray}}
\newcommand{\eeqa}{\end{eqnarray}}

\def \del {\partial}
\def \K {{\cal K}}
\def \mn {{\mu\nu}}
\def \L {\Lambda_3}

\newcommand{\order}{{\cal O}}

\begin{document}

\title{Strong Coupling and Bounds on the Graviton Mass in Massive Gravity}

\author{Clare Burrage}
\affiliation{School of Physics and Astronomy, 
University of Nottingham, Nottingham NG7 2RD, UK} 
\author{Nemanja Kaloper} 
\affiliation{Department of Physics, University of California, Davis, CA95616, USA} 
\author{Antonio Padilla} 
\affiliation{School of Physics and Astronomy, 
University of Nottingham, Nottingham NG7 2RD, UK}

\date{\today}

\begin{abstract}
The theory of a single massive graviton has a cutoff much below its Planck scale, because the extra modes from the graviton
multiplet involve higher derivative self-interactions, controlled by a scale convoluted from the small graviton mass.
On a generic background, these correct the propagator by environmental effects. 
The resulting effective cutoff depends on the environmental parameters and the graviton mass. Requiring the theory to be perturbative down to ${\cal O}(1)\mbox{ mm}$, we derive bounds on the graviton mass, corresponding to $\gtrsim {\cal O}(1) \mbox{ meV}$ for the generic case, and somewhat weaker bounds in cases of fine-tuning. In all cases the mass is required to be much too large for the theory to conform with GR at cosmological distances. Similar results are also found in quartic and quintic Galileon theory.
\end{abstract}

\maketitle

What is the range of the gravitational force? Must it be infinite, or could it be finite, by virtue of a graviton mass-induced Yukawa suppression,
like in massive gauge theories? This question has been looming about a long time, since the pioneering work
by Pauli and Fierz \cite{pf}, the subsequent exploration by Boulware and Deser \cite{bdghost}, and its recent followup \cite{crenictrin}. The problem one encounters is that since mass breaks the residual gauge symmetries of gravity, there are six new propagating degrees of freedom. Generically, one is a ghost. While Pauli and Fierz exorcised the ghost away in the linearized limit, it seemed unavoidable in the full theory \cite{bdghost,crenictrin}. On the other hand, discovery of cosmic acceleration \cite{sn} and the dearth of its theoretical explanations, save the landscape paradigm \cite{weinberg} (many less satisfactory dark energies are reviewed in
\cite{dereview}) fueled speculations that changing gravity away from General Relativity (GR) may account for dark energy \cite{mgreview}. For example, this could happen if one could give the graviton a mass of order the current Hubble scale $H_0 \sim 10^{-33}$ eV. Hence the question: {\it can the graviton have a mass?} becomes more than just a mere theoretical curiosity.

Construction of classically consistent massive gravity has been difficult (for a review see \cite{kurtreview}). Linearised GR with a Pauli-Fierz (PF) mass term \cite{pf} suffers from the vDVZ discontinuity \cite{vdvz}, and its linearized perturbation theory is unreliable.
This can be improved by non-linear interactions and the implementation of the Vainshtein mechanism \cite{vain,dysphere}. However, typical non-linear completions have the sixth mode Boulware-Deser ghost \cite{bdghost}, in addition to the five standard massive (and healthy) helicities of Poincare-invariant spin-2 theory. Most of these issues are closely related to the  dynamics of the helicity-0 component of the massive graviton, 
hereon denoted $\pi$.

Very recently, it has been shown that many problems can be avoided in a specific non-linear completion of PF theory, known as dRGT \cite{genpf,resum}.
Classically, the dRGT model does not propagate the troublesome sixth mode \cite{noghost}.  Hence, it gives a fully self-consistent classical system with massive spin-2 that can be used as a straw-man for phenomenological purposes, such as studying the experimental bounds on the mass of the graviton, with a definable perturbative expansion at any order of truncation of the theory. However, the full taxonomy of the background solutions on which to expand still does not exist (for some problems, see \cite{guido,shinji}).

In this {\it Letter} we will address the phenomenological limits on the graviton mass. Since the theory has a UV cutoff much below the Planck scale, 
if one wishes to use it to approximate GR one must require it to be perturbatively well behaved at least between the distances scales of ${\cal O}(1) $ mm, and the present Hubble scale, the range where we have more or less found gravity to be weak. This places a bound on the mass of the graviton, which is directly related to the UV cutoff. To determine it quantitatively, one must include environmental effects in the calculation of the short distance cutoff, which become important due to the higher derivative self interactions of the extra scalar mode $\pi$ in the graviton spectrum. 

We will infer the bounds as follows. Working at distances shorter
that the inverse graviton mass, we can ignore the Yukawa suppressions, and focus on the dynamics of the Stuckelberg field  $\pi$, being guided by the Goldstone equivalence theorem from massive gauge theory. This is justified because in massive gravity the (low) UV cutoff is (still) higher than the mass of the graviton, and so there is a (broad) regime of  scales where this approximation is valid. We will ignore the dynamics of the helicity-1 modes, since by Lorentz invariance they are subleading at the lowest order. We will then compute the effective action in the background fields of the Earth, a framework appropriate for comparisons with the results of the tabletop experiments that probe gravity on the smallest scales \cite{adelberger}, and extract the strong coupling scale for it. As we will then see, making the strong coupling scale high enough will drag up the graviton mass much above the current Hubble scale.

\paragraph{Framework.} For a graviton of mass $m$, the dRGT theory is described by the following action \cite{genpf,resum}
\be \label{act}
S=\frac{1}{\kappa^{2}} \int d^4 x  \sqrt{-g} \left[R -\frac{m^2}{4} {\cal U}(g, H)\right]+S_m[g_\mn; \Psi_n]
\ee
with $\kappa=\sqrt{16\pi G} =\sqrt{2}/M_{pl}$ and we have graviton potential ${\cal U}$. $S_m$ is the action for matter fields, $\Psi_n$, minimally coupled to the metric $g_\mn$. The covariant tensor $H_\mn$ is related to the metric as $g_\mn=\eta_\mn+\kappa h_\mn=H_\mn+\eta_{ab} \del_\mu \phi^a \del_\nu \phi^b$
where the four Stuckelberg fields, $\phi^a$ transform as scalars and $\eta_{ab}=\textrm{diag}(-1,1,1,1)$.  The potential can be expressed using $\K^\mu_\nu=\delta^\mu_\nu-\sqrt{\delta^\mu_\nu-H^\mu_\nu}$, so that we have 
${\cal U}(g, H)=-4\sum_{n \geq 2} \alpha_n \K^{[\mu_1}_{\nu_1} \ldots  \K^{\mu_n]}_{\nu_n}$,. The square brackets denote antisymmetrization, {\it without} the usual factor of $1/n!$. We can extract the helicity-0 component $\pi$ of the graviton by setting $\phi^a=x^a-\eta^{a \mu} \del_\mu \pi$.

To zoom in on the interesting dynamics, one usually takes the decoupling limit \cite{genpf}: $m, \kappa \to 0, ~T_\mn \to \infty$ with
$\L=(m^2/ \kappa)^{1/3}$ and $ \kappa T_\mn$ held fixed, where $T_\mn=-\frac{2}{\sqrt{-g}} \frac{\delta S_m}{\delta g^\mn}$ is the energy-momentum tensor of the source.  The effective Lagrangian in this limit is \cite{genpf}
\begin{multline} \label{lag}
{\cal L}=-\frac{1}{2} h_\mn {\cal E}^{\mn \alpha \beta} h_{\alpha \beta} +\frac{3}{2} \pi \Box \pi \\-\frac{u}{ \L^3} \pi \pi^{[\mu}_\mu \pi^{\nu]}_\nu+\frac{1}{4\L^6} (u^2-4v) \pi \pi^{[\mu}_\mu \pi^\nu_\nu \pi^{\alpha]}_\alpha  \\+\frac{3v}{\L^6}\left (h_\mn -\frac{1}{3} h^\gamma_\gamma \eta_\mn \right) \pi^{\mu[\nu} \pi^\alpha_\alpha \pi^{\beta]}_\beta+\frac{uv}{\L^9} \pi \pi \pi^{[\mu}_\mu \pi^\nu_\nu \pi^{\alpha}_\alpha \pi^{\beta]}_\beta \\
+\frac{\kappa }{2}  h^\mn T_\mn +\frac{\kappa }{2} \pi T^\alpha_\alpha+\frac{\kappa u}{2 \L^3} \del^\mu  \pi \del^\nu \pi T_\mn
\end{multline}
where $u=-(1+3\alpha_3)$, $v=-\frac{1}{2} (\alpha_3+4\alpha_4)$, $\pi_\mn=\del_\mu \del_\nu \pi$ and indices are raised/lowered with the fiducial  Minkowski metric. The operator ${\cal E}^{\mn \alpha \beta} $ is related to the linearised Einstein tensor\footnote{${\cal E}^{\mn \alpha \beta} h_{\alpha \beta}=\delta G^\mn=-\frac{1}{2} \Box \left(h^\mn-\frac{1}{2} h^\alpha_\alpha \eta^\mn \right)+\ldots$}. Note that we have performed the following field redefinitions:  $\pi \to \pi/\L^3$,
$h_\mn \to h_\mn+\pi \eta_\mn+\frac{u}{\L^3} \del_\mu \pi \del_\nu \pi$, the latter to diagonalise the action up to  cubic  order. It is impossible to fully diagonalise the theory in an explicitly local way. 

Clearly, the interactions in (\ref{lag}) become strongly coupled at the scale $\Lambda_3$. For a graviton whose mass lies at the current Hubble scale, $m \sim H_0 \sim 10^{-33}$ eV, in vacuum this occurs at distances $\la 1000$ km \cite{nicrattazzi}. 
However, in the presence of the Earth's background fields, the quadratic Lagrangian in (\ref{lag}) will be renormalized by the contributions from the higher dimension operators in (\ref{lag}) evaluated on the background. To compute them, we model the Earth's background field with the spherically symmetric static solutions found in \cite{gus}. In the decoupling limit this is given by
\be \label{earthfield}
ds^2=(-1+\kappa \bar h_{tt}) dt^2+(1+\kappa \bar h_{\rho\rho}) d\rho^2 +\rho^2 d\Omega^2
\ee
where $ \bar h_{tt}(\rho)=\int^\rho dz \frac{\bar h_{\rho\rho}(z)}{z}$, $\bar h_{\rho\rho}(\rho)=\frac{\kappa M}{8 \pi \rho}+2v \rho^2\L^3 Q^3$,
and $Q=\frac{\bar \pi'(\rho)}{ \L^3 \rho}$ satisfies
\be \label{Qeq}
3Q-6uQ^2+2(u^2-4v) Q^3-6vQ^2 \left(\frac{ \bar h_{\rho\rho}}{\L^3 \rho^2} \right)=\frac{1}{4\pi} \left(\frac{\rho_V}{\rho} \right)^3
\ee
Here $M$ is the mass of the Earth, and $\rho_V=\frac{(\kappa M)^\frac{1}{3}}{\L}$ its Vainshtein radius.  
Requiring that the Vainshtein shielding is efficient, such that that $|\bar \pi| \ll |\bar h_{\mu\nu}|$ for $\rho \ll \rho_V$, for generic values 
$|u|, |v| \sim \order (1)$, we must require $v<0$ \cite{gus}, in which case we obtain $Q \sim \order (1)$ and  $|\bar h_{\mu\nu}| \sim \order \left( \frac{\kappa M}{8 \pi \rho}\right)$, implying  $|\bar \pi| \ll |\bar h_{\mu\nu}|$, as desired.  We see similar behavior when $|u| \ll 1$ and $|v| \sim \order (1)$. For $|u| \sim \order (1)$ and $|v| \ll 1$ we have $Q \sim {\cal O}(\frac{\rho_V}{\rho})$ and $|\bar h_{\mu\nu}| \sim {\cal O}( \frac{\kappa M}{8\pi  \rho})$, so again, the Vainshtein mechanism is successful. 

The Lagrangian (\ref{lag}) omits irrelevant operators suppressed by scales between $\Lambda_3$ and the Planck scale. In fact,  one can easily check  using the exact results of \cite{gus} that these operators are subleading,  and remain subdominant for the processes considered here. Eq. (\ref{lag}) is a sufficiently good approximation to the full theory for momenta in the range $ m \ll p \ll \tilde \Lambda_\oplus$, with $\tilde \Lambda_\oplus$ the running strong coupling scale on Earth \cite{nicrattazzi}. 
\paragraph{Effective theory.} Let us now determine the effective theory in the background field of the Earth. This means, we perturb about the background solution (\ref{earthfield}), setting $h_\mn=\bar h_\mn+\chi_\mn$, $\pi=\bar \pi+\varphi$ and $T_\mn=\bar T_\mn+\tau_\mn$. Working with the  Lagrangian (\ref{lag}) and defining
$[r, s]_\mn = r \varphi_{\mu[\nu} \varphi^{\mu_2}_{\mu_2} \ldots  \varphi^{\mu_r}_{\mu_r}\bar \pi^{\nu_1}_{\nu_1} \ldots \bar \pi^{\nu_s}_{\nu_s]}+s\bar \pi_{\mu[\nu} \varphi^{\mu_1}_{\mu_1} \ldots  \varphi^{\mu_r}_{\mu_r}\bar \pi^{\nu_2}_{\nu_2} \ldots \bar \pi^{\nu_s}_{\nu_s]}$
 and $
[r, s] =\varphi^{\mu_1}_{[\mu_1} \ldots  \varphi^{\mu_r}_{\mu_r}\bar \pi^{\nu_1}_{\nu_1} \ldots \bar \pi^{\nu_s}_{\nu_s]} $,
 we obtain
\begin{multline}
\delta {\cal L}=-\frac{1}{2} \chi_\mn {\cal E}^{\mn \alpha \beta} \chi_{\alpha\beta}+\frac{1}{2}\varphi {\cal K} \varphi\\+\frac{3v}{\L^6} \chi^\mn ([1,2]_\mn-\eta_\mn [1,2])+\delta {\cal L}_{int} +\delta {\cal L}_{m} 
\end{multline}
where
\ba
&&\varphi {\cal K} \varphi = \frac{6v}{\L^6}\bar h^\mn([2,1]_\mn-\eta_\mn [2,1]) + \varphi\Bigl[3 \Box \varphi  
-\frac{6u}{\L^3} [1,1] \nonumber \\
&&~~~~ +\frac{3(u^2-4v)}{\L^6} [1,2] +\frac{20uv}{\L^9} [1,3]
-\frac{\kappa u}{ \L^3} \bar T^\mn \varphi_\mn\Bigr] \nonumber \\
&&\delta {\cal L}_{int} =-\frac{u}{\L^3} \varphi[2, 0]+\frac{1}{4 \L^6} (u^2-4v) \varphi(4[2,1]+[3,0]) \nonumber\\
&&~~~~ +\frac{uv}{\L^9} \varphi (10[2,2]+5[3,1]+[4,0]) \nonumber \\
&&~~~~ +\frac{v}{\L^6} \left(\bar h^\mn-\frac{1}{3} \bar h^\alpha_\alpha \eta^\mn\right)[3,0]_\mn \\
&&~~~~ +\frac{v}{\L^6} \left(\chi^\mn-\frac{1}{3} \chi^\alpha_\alpha \eta^\mn\right) (3[2,1]_\mn+[3,0]_\mn) \nonumber \\
&&\delta {\cal L}_m = \frac{\kappa}{2}\chi^\mn \tau_\mn+\frac{\kappa}{2} \varphi \tau^\alpha_\alpha 
 +\frac{\kappa u}{\L^3} \varphi \bar \pi^\mn \tau_\mn-\frac{\kappa u}{2\L^3}   \varphi \varphi^\mn \tau_\mn \nonumber
\ea
Next, we diagonalize the bilinears by means of the non-local field redefinition $\chi_\mn =\tilde  \chi_\mn+3v A_\mn$, where $A_\mn=-\frac{2}{\L^6}\Box^{-1}\left([1,2]_\mn-\frac{1}{2}\eta_\mn[1,2]\right)$. First, we obtain:
\begin{multline}
\delta {\cal L}=-\frac{1}{2} \tilde \chi_\mn {\cal E}^{\mn \alpha \beta} \tilde \chi_{\alpha\beta}+\frac{1}{2}\varphi {\cal K} \varphi \\+\frac{9v^2}{\L^6}  A^\mn\left([1,2]_\mn-\eta_\mn[1,2]\right)
+\delta {\cal L}_{int} +\delta {\cal L}_{m} 
\end{multline}
In what follows we neglect the non-local contribution to the scalar propagator. This is legitimate because the nonlocal terms are systematically smaller from the contributions $\propto \bar h_{\mu\nu}$ inside the Vainshtein radius, as can be directly checked.
 Similarly, the change of variables introduces a new coupling to matter of the form $\frac{3}{2} v \kappa A^\mn \tau_\mn \lesssim \frac{\kappa}{2} \varphi \tau^\alpha_\alpha $,  which we also neglect since $r, \rho \lesssim \rho_V$. 
 
\paragraph{The strong coupling scale}
To find the effective strong coupling scale in Earth's background, $\Lambda_\oplus$, note that the linearised fluctuations are described by
\be
\delta {\cal L}_{kin} =-\frac{1}{2} \tilde \chi_\mn {\cal E}^{\mn \alpha \beta} \tilde \chi_{\alpha\beta}-\frac{1}{2} \varphi (\xi \del_t^2 -P^{ij} \del_i \del_j)\varphi
\ee
where 
\ba
&&\xi = 3 -\frac{6u}{\L^3}[0,1]+\frac{3}{\L^6} (u^2-4v)[0,2]+\frac{20uv}{\L^9}[0,3]  \nonumber \\
&&~~~~ +\frac{6v}{\L^6}\left[ (\hat h^{t}_t\bar \pi^{[i}_i)^{j]}_j   + (\bar \pi^k_i \hat h^{[i}_k)^{j]}_j  +(\bar \pi^{[i}_i \hat h^{j]}_k)^{k}_j \right] \nonumber \\
&& P^{ij} k_i k_j  = 3 |\uvec{k} |^2-\frac{6u}{\L^3} D_{[1,1]}(\k)+\frac{3}{\L^6} (u^2-4v)D_{[1,2]}(\k)\nonumber \\
&&~~~~ +\frac{20uv}{\L^9} D_{[1,3]} (\k) +\frac{6v}{\L^6}\left[   (\hat h^{[i}_j\bar \pi^k_k)^{l]}_l  \k_i \k^j \right. \nonumber \\
&&\left.~~~~ + (\bar \pi^l_i \hat h^{[i}_l)^{j}_j \k^{k]} \k_k+ \k^{[i} \k_i(\bar \pi^{j}_j \hat h^{k]}_l)^{l}_k \right]
\ea
and $D_{[1, s]}(\k)=\k^i \k_{[i} \bar \pi^{j_1}_{j_1} \ldots \bar \pi^{j_s}_{j_s]}$, $\hat h_\mn=\bar h_\mn-\frac{1}{3}\bar h^\alpha_\alpha \eta_\mn$. 
We neglect the non-local contributions and use $\bar T_\mn=0$ outside the source
.  The interactions, schematically, are
\be
{\cal I}= \frac{f(u, v)(\bar h)^{a}(\tilde \chi)^{b}\varphi^{c}(\Box^{-1})^d [\alpha, \beta]
}{\Lambda_3^{a+b+c-2d+3\alpha+3\beta-4}} \, . \ee
Now, to extract the strong coupling scale(s), we first canonically normalise the kinetic term. Noting
$\xi, P \sim \order(1)+u\order\left(\frac{\del \del \bar \pi}{\L^3}\right)+(u^2-4v)\order\left(\frac{\del \del \bar \pi}{\L^3}\right)^2+uv \order\left(\frac{\del \del \bar \pi}{\L^3}\right)^3+v\order\left( \frac{\bar h \del \del \bar \pi}{\rho^2\L^6}\right)$, when $|u|, |v| \sim \order(1)$, the last term dominates inside the Vainshtein radius and so $\xi \sim P \sim \left(\frac{\rho_V}{\rho}\right)^3 \gg 1$.  The same is true when $|u| \ll 1, |v| \sim \order(1)$.  In contrast, when $|u| \sim \order(1)$, $|v| \ll 1$,
$\bar \pi \propto \rho$ inside the Vainshtein radius, and so $\bar \pi''(\rho) \ll \bar \pi'(\rho)/\rho$ \cite{clare}, which leads to a hierarchy of eigenvalues for $P$.  In particular, we find one very large eigenvalue $P_{1} \sim \xi \sim Q^2 \sim \left(\frac{\rho_V}{\rho}\right)^2$, and two smaller eigenvalues $P_2 \sim P_3 \sim Q\sim \frac{\rho_V}{\rho}$

Thus, for $|u|, |v| \sim \order(1)$ and $|u| \ll 1, |v| \sim \order(1)$, the canonical scalar field is $\hat \varphi \sim \left(\frac{\rho_V}{\rho} \right)^{3/2} \varphi$. Using $\bar h \sim {\cal O}\left( \frac{\kappa M}{8\pi  \rho}\right)$, and $\del \del \pi \sim \Lambda_3^3 Q \sim \Lambda_3^3$, an interaction
\be
{\cal I}= \frac{f(u, v)\left( \frac{\kappa M}{8\pi  \rho}\right)^{a}\left(\frac{\rho_V}{\rho}\right)^{-\frac{3}{2}(c+\alpha)}(\tilde \chi)^{b}\hat \varphi^{c}(\Box^{-1})^d (\del \del \hat\varphi)^\alpha 
}{\Lambda_3^{a+b+c-2d+3\alpha-4}}\ee
becomes strong at the scale
\be
\Lambda_{\cal I} \sim \Lambda_3 \left[\frac{(8\pi)^a (\rho \Lambda_3)^{a-\frac{3}{2}(c+\alpha)}}{f (u, v)(\kappa M)^{a-\frac{1}{2}(c+\alpha)}}\right]^\frac{1}{b+c-2d+3\alpha-4} \, .
\ee
The theory clearly becomes nonperturbative at the lowest such scale coming from any interactions present. It turns out that the lowest strong coupling scale arises from  $\frac{v}{\L^6} \left(\bar h^\mn-\frac{1}{3} \bar h^\alpha_\alpha \eta^\mn\right)[3,0]_\mn $, and gives an energy-momentum cut-off
\be
\Lambda_\oplus \sim \frac{1}{km} \left(\frac{m}{H_0}\right)^{1/5} \, .
\ee
Here we use $\rho \sim 6000$ km and $M \sim 10^{33}/\kappa$ for Earth's radius and mass, respectively. So for a graviton with Hubble mass, the theory is strongly coupled below the kilometer scale! This differs from DGP on Earth \cite{nicrattazzi} due to the graviton-scalar mixing in (\ref{lag}), which enhances the scalar three-point vertex by the background gravitational field, that dominates due to the Vainshtein effect. 
To conform with tabletop experiments \cite{adelberger}, we must suppress 
$\Lambda_\oplus$ by six orders of magnitude, to ${\cal O}(1)$ mm. This pushes the graviton mass up by $\ga 30$ orders of magnitude, to 
$m \ga 10^{-3}$ eV. So heavy a graviton would experience Yukawa suppression at distances longer that around a millimeter, failing to conform with GR at all currently macroscopically tested scales \cite{adelberger}.

One might hope that in the special limits $|u| \sim \order(1)$, $|v| \ll 1$ the theory is much better behaved because the menacing graviton-scalar mixing is absent when $v=0$. However,  this is not quite so; while the bounds on the graviton mass are weaker, they are still significant. First, recall that now there is a hierarchy of eigenvalues for $P$. Using an orthogonal coordinate transformation to diagonalize $P$, the kinetic term  for the scalar is given by
\be
\delta {\cal L}_{kin} \supset -\frac{1}{2} \varphi (\xi \del_t^2 -P_1 \del_1^2-P_2 \del_2^2-P_3 \del_3^2 )\varphi \, .
\ee
We canonically normalise it by  stretching two of the space directions, $( \hat t, \hat x_1, \hat x_2, \hat x_3)=\left(t, x_1, x_2\sqrt{\frac{\rho_V}{\rho}}, x_3\sqrt{\frac{\rho_V}{\rho}}\right) $, and defining $\hat \varphi \sim\varphi \sqrt{\frac{\rho_V}{\rho}}$, which yields the interactions
\be
d^4 x \frac{\varphi [\alpha, \beta]}{\Lambda_3^{3(\alpha+\beta-1)}} \sim  d^4 x \left(\frac{\rho_V}{\rho}\right)^{\beta-\frac{1}{2}(1+\alpha)} \frac{\hat \varphi (\del \del \hat\varphi)^\alpha}{\Lambda_3^{3(\alpha-1)}} \, ,
\ee
where we have used the fact that $\del \del \pi \sim \Lambda_3^3 Q \sim \Lambda_3^3\left(\frac{\rho_V}{\rho}\right)$ for the fine-tuned scenario. Given  that table top gravity experiments probe spatial rather than temporal fluctuations in the gravitational field, we do not consider interactions with temporal derivatives. The largest interaction comes from the term
\begin{multline} \label{largeints}
d^4 x \frac{u^2-4v}{\Lambda_3^6} \varphi [2, 1] \supset d^4 \hat x \left[ \frac{\order(1)}{\Lambda_3^3}\left(\frac{\rho_V}{\rho}\right)^{-\frac{1}{2}}\hat\varphi \hat \del_1^{2} \hat \varphi\hat \del_\perp ^{2} \hat \varphi \right.\\\left.  + \frac{\order(1)}{\Lambda_3^3}\left(\frac{\rho_V}{\rho}\right)^{\frac{1}{2}}\hat \varphi (\hat \del_\perp ^{2} \hat \varphi)^2 \right]
\end{multline}
where we  are being somewhat schematic on the RHS and have explicitly switched to the stretched coordinate system. As in \cite{gal}, we have introduced a shorthand $\perp=2, 3$.  Note the absence of an interaction of the schematic form $\hat \varphi (\hat \del_1^2 \hat \varphi)^2$ which cannot exist by virtue of the antisymmetrisation\footnote{This fact was not entirely appreciated in \cite{gal}}. Following \cite{gal}, an analysis of the $2 \to 2$ scattering amplitude including these interactions reveals a physical momentum cut-off
\be \label{scscale2}
\Lambda_\oplus^{(k)} \sim \frac{1}{20 km} \left(\frac{m}{H_0}\right)^{5/9}
 \ee
in the Earth's background field, where we have taken extra care to include the appropriate recaling when switching back from stretched to physical coordinates\footnote{Note, however, that the lowest strong coupling scale arises from the first term in Eq. \ref{largeints}, and is given along the unscaled radial direction.}. So in this case a Hubble mass graviton leads to a breakdown of predictability due to quantum effects on scales of tens of kilometres. The reason this differs considerably from centimeter scale in DGP on Earth \cite{nicrattazzi} is that in DGP only terms up to cubic order in the scalar are considered. Our strong coupling scale derives from the term that is quartic in the original $\pi$ field. We also need the quartic interaction to pick up the hierarchy in the eigenvalues of the background matrix $P$. So, to push the strong coupling scale in this limit down to the a millimeter we must require $m \gtrsim 10^{-15}$ eV. This places the Vainshtein radius of the Sun at $\lesssim  10^4$ km, well inside the orbit of Mercury, and also implying that outside of the Solar System the full potential has Yukawa suppression, in contrast with GR.

\paragraph{Strong coupling in Galileon theory} The finely tuned limit above corresponds to precisely a quartic Galileon theory in the decoupling limit \cite{gal}, generically described by a Lagrangian of the form
\be
{\cal L}=\frac{3}{2} \pi \Box \pi+\sum_{n=2}^4 \frac{c_n}{\Lambda^{3(n-1)}} \pi \pi^{\mu_1}_{[\mu_1} \cdots \pi \pi^{\mu_n}_{\mu_n]}+\frac{\kappa}{2} \pi T
\ee
where the $c_n$ are constants $\sim \order(1)$, and the scale $\Lambda$ is usually related to the Hubble scale such that $\Lambda \sim (H_0^2/\kappa)^{1/3} \sim 1/1000km$ \cite{gal}. So our results from this limit extend to Galileons too.

For the cubic Galileon theory ($c_3=c_4=0$), the analysis of \cite{nicrattazzi} holds and one gets a running strong coupling scale in the Earth's gravitational field of $\Lambda_\oplus \sim \frac{1}{1cm}(\Lambda \cdot 1000km)^{1/4}  $, giving the well known centimeter scale in the usual case.  For the quartic Galileon ($c_4=0$), the analysis of the fine-tuned case of the previous section holds, provided we identify our $\Lambda $ with $\Lambda_3$, giving the effective strong coupling scale, $\Lambda_\oplus \sim \frac{1}{20 km}(\Lambda \cdot 1000km)^{5/6} $.

In the case of the quintic Galileon, the background field is still $\bar \pi =\Lambda^3 \rho_V \rho$, where the Vainshtein radius is $\rho_V \sim (\kappa M)^{1/3}/\Lambda$ \cite{clare}. The analogue of the matrix $P$ also coincides with the quartic case, having eigenvalues $P_1 \sim\left(\frac{\rho_V}{\rho}\right)^2, P_2 \sim P_3 \sim  \frac{\rho_V}{\rho}$.  An important difference is that $\xi \sim \left(\frac{\rho_V}{\rho}\right)^3$, which helps to raise the cut-off slightly. The rescaling to canonical coordinates is $( \hat t, \hat x_1, \hat x_2, \hat x_3)=\left(t, x_1\sqrt{\frac{\rho_V}{\rho}}, x_2\frac{\rho_V}{\rho}, x_3\frac{\rho_V}{\rho}\right) $, and the canonical field is $\hat \varphi \sim \left(\frac{\rho_V}{\rho}\right)^{1/4} \varphi$. Again, the largest interactions come from terms analogous to (\ref{largeints}), this time giving rise to a running strong coupling scale of $\Lambda_\oplus \sim \frac{1}{50 \text{metres}}(\Lambda \cdot 1000km)^{7/12} $.

\paragraph{Summary} We have shown that massive gravity generically suffers from serious strong coupling problems and a loss of predictivity at unacceptably low scales,  which can be avoided only by requiring the graviton mass to be much higher than the present Hubble scale. This implies that the theory will not approximate GR at cosmological scales. We note similar behavior in quartic and quintic Galileon theory, noted previously in \cite{gal}: if the quartic and/or quintic terms are significant enough to affect the cosmological background today then we also find loss of predictivity at unacceptably low scales rendering the model phenomenologically inept.

{\bf Acknowledgments}:~We thank G. D'Amico and  G. Tasinato for useful discussions.
NK thanks the School of Physics and Astronomy, Univ. of Nottingham, for kind hospitality in the course of this work.
CB was funded by a Univ. of Nottingham Anne McLaren fellowship. NK is supported by the DOE Grant DE-FG03-91ER40674.
AP was funded by a Royal Society URF. 

\appendix

\section*{Including a background matter profile}

Throughout this paper we have explicitly assumed that the background energy momentum tensor vanishes $\bar T_{\mu\nu}=0$. It has been suggested \cite{nonren} that this assumption is crucial to our results, and that the explicit inclusion of the background matter profile due to the earth and its atmosphere will alter our findings. Here we will show that our bounds remain completely unaltered. To this end, let us assume that the background energy momentum tensor, $\bar T_{\mu\nu}=\text{diag}(\epsilon(\rho),0, 0, 0)$, where we have neglected pressure, and take a spherically symmetric energy density profile
\be
\epsilon(\rho)=\begin{cases} \epsilon_\oplus & \rho< \rho_\oplus \text{~"inside the earth"} \\
\epsilon_{atm} & \rho_\oplus <\rho< \rho_{atm}  \text{~"inside the atmosphere"} \\
0 & \rho>\rho_{atm}  \text{~"outer-space"} \end{cases}
\ee  
where $\rho_\oplus \sim 6000km $ and $\rho_{atm} \sim 1.01 \rho_\oplus$ are the radii of the earth and atmosphere respectively. The energy density of the earth is given by $\epsilon_\oplus = M/(4\pi \rho_\oplus^3/3)$ and of the atmosphere by $\epsilon_{atm}\approx 10^{-3} \epsilon_\oplus$.  The mass contained with a radius $\rho$ is easily computed, and is given by
\be
M(\rho)=\begin{cases} M\rho^3/\rho_\oplus^3  & \rho< \rho_\oplus\\
M \left(1+ (\rho^3/\rho_\oplus^3-1)\delta\right)& \rho_\oplus <\rho< \rho_{atm}  \\
M \left(1+ (\rho_{atm}^3/\rho_\oplus^3-1)\delta\right) & \rho>\rho_{atm}  \end{cases}
\ee  
where $\delta=\epsilon_{atm}/\epsilon_\oplus \sim 10^{-3}$. In the absence of pressure, we still have $ \bar h_{tt}(\rho)=\int^\rho dz \frac{\bar h_{\rho\rho}(z)}{z}$,  but with $\bar h_{\rho\rho}(\rho)=\frac{\kappa M(\rho)}{8 \pi \rho}+2v \rho^2\L^3 Q^3$,
and $Q=\frac{\bar \pi'(\rho)}{ \L^3 \rho}$ satisfying
\be \label{newQeq}
3Q-6uQ^2+2(u^2-4v) Q^3-6vQ^2 \left(\frac{ \bar h_{\rho\rho}}{\L^3 \rho^2} \right)=\frac{1}{4\pi}\frac{\kappa M(\rho)}{\Lambda_3^3 \rho^3} 
\ee
Let us first consider the generic situation, for which $|u| \lesssim \order(1)$ and $|v| \sim \order(1)$. As before, we find that $Q \sim 1$, and $\bar h_{\rho\rho} \sim \frac{\kappa M(\rho)}{8\pi \rho}$. As regards fluctuations, it is easy to see that $\xi$ receives an additional contribution that goes like $\Delta \xi =\frac{\kappa u \epsilon}{\Lambda^3_3}$.  Previously, $\xi$ was dominated by the term $v\order\left( \frac{\bar h \del \del \bar \pi}{\rho^2\L^6}\right) \sim \frac{\kappa M(\rho)}{\Lambda_3^3 \rho^3}$. In the atmosphere where table top gravity experiments are typically conducted, we take $\rho =\lambda \rho_\oplus$, where $\lambda \in [1, 1.01]$ and note that  $\Delta \xi  \sim \delta \left(\frac{\rho_V}{\rho_\oplus }\right)^3$ while  $v\order\left( \frac{\bar h \del \del \bar \pi}{\rho^2\L^6}\right) \sim  \left(\frac{\rho_V}{\rho_\oplus }\right)^3(\lambda^{-3} +(1-\lambda^{-3}) \delta) \gg \Delta \xi$ since $\delta \ll 1$.  We conclude that, as before, we have $\xi \sim \left(\frac{\rho_V}{\rho_\oplus }\right)^3$ to a good approximation.  In the absence of pressure, we find that $P_{ij}$ receives no additional contributions, and one can similarly argue that we once again have $P  \sim \left(\frac{\rho_V}{\rho_\oplus }\right)^3$. We therefore conclude that in the generic case ($|u| \lesssim \order(1)$ and $|v| \sim \order(1)$), the analysis of the main text should hold to a good approximation when applied to experimental gravity tests conducted inside the atmosphere.

The situation is a little more subtle in the fine -tuned scenario $|u| \sim \order(1)$ and $|v| \ll 1$.  We now have $Q \sim \frac{\kappa M(\rho))^{1/3}}{\Lambda_3 \rho}$ and $\bar h_{\rho\rho} \sim \frac{\kappa M(\rho)}{8\pi \rho}$.  In particular, in the atmosphere at $\rho =\lambda \rho_\oplus$ where   $\lambda \in [1, 1.01]$ we find $Q \sim \frac{\rho_V}{\rho_\oplus} \left[\lambda^{-3} +(1-\lambda^{-3}) \delta\right]^{1/3}$. It is not immediately clear whether the additional contribution $\Delta \xi$ is significant so for the moment we simply include it alongside the dominant contribution $\sim Q^2$ from the remainder of $\xi$. In the atmosphere, we therefore obtain
\be
\xi \sim Q^2 +\order (1) \frac{\kappa \epsilon_{atm}}{\Lambda_3^3} \sim Q^3 \left[ \frac{1}{Q}+ \order(1) \delta \right]
\ee
As regards $P_{ij}$ we once again find a hierarchy of eigenvalues. This follows from the fact that
\be
\frac{\bar \pi''}{(\bar \pi'/\rho)} \sim \frac{\rho M'(\rho)}{3M( \rho)}\sim \order(1) \delta \qquad \text{at $ \rho= \lambda \rho_\oplus$}
\ee
such that we have $P_1 \sim Q^2$ and $P_2 \sim P_3 \sim Q^2  \left[ \frac{1}{Q}+ \order(1) \delta \right]$. To canonically normalise the kinetic term, we introduce new coordinates $(\hat t, \hat x_1, \hat x_2,  \hat x_3)=(t, Ax_1, Bx_2, Bx_3)$ and $\hat \varphi=\frac{\sqrt{\xi}}{B\sqrt{A}} \varphi$, where $A=\sqrt{\frac{\xi}{P_1}}$ and $B=\sqrt{\frac{\xi}{P_2}}$. Focussing on the same interaction as the the one identified previously
$$
\frac{1}{\Lambda_3^6} \varphi [2,1] \supset \frac{\order(1)}{\Lambda_3^6} \varphi \del_1^2 \varphi \del_\perp^2 \varphi\del_\perp^2 \bar \pi+\frac{\order(1)}{\Lambda_3^6} \varphi( \del_\perp^2\varphi )^2 \del_1^2 \bar \pi
$$
we identify a physical momentum cut-off
\be
\Lambda^{(k)}_\oplus \sim \Lambda_3 Q^{3/4}  \left[ \frac{1}{Q}+ \order(1) \delta \right]^{7/12}
\ee
where we recall that everything has been evaluated in the atmosphere where $\rho =\lambda \rho_\oplus$ and   $\lambda \in [1, 1.01]$, and we have done the appropriate rescaling when switching from stretched to physical coordinates.  Now if $m \gtrsim 10^{15} \delta^{3/2} H_0$, it turns out that $\frac{1}{Q} \gtrsim \delta$ and we recover the result of our previous analysis given by equation (\ref{scscale2}). In contrast, if $m \lesssim 10^{15} \delta^{3/2} H_0$, then $\frac{1}{Q} \lesssim \delta$ and we obtain
\be
\Lambda^{(k)}_\oplus \sim \frac{1}{2 cm}\delta^{7/12} \left(\frac{m}{H_0}\right)^{1/6} \sim \frac{1}{1.2m} \left(\frac{m}{H_0}\right)^{1/6}
\ee
This suggests a cut-off of order $1.2m$ in the fine-tuned scenario when the graviton has Hubble mass. While this is an improvement on the previous estimate, it is still many orders of magnitude too large.  In addition, the scaling with the graviton mass has become much weaker which means one must increase its value considerably in order to make the strong coupling scale tolerable. To push strong coupling down to the millimeter scale one might infer that $m \gtrsim 10^{-15}$ eV, which is actually the same bound as the one obtained previously.
%

Finally, we note that in reality torsion balance experiments are performed in a near vacuum, with $\epsilon_{experiment} \sim 10^{-14} \epsilon_{atm}$. In other words we have $\delta_{experiment} \sim 10^{-17}$, rendering the explicit inclusion of matter somewhat nugatory.

\end{document}